\documentclass[draftcls,onecolumn]{IEEEtran}
\usepackage{upgreek}
\usepackage{graphicx}
\usepackage{amsmath}
\usepackage{multicol}
\usepackage{multirow}
\usepackage{stfloats}
\usepackage{booktabs}
\usepackage{mathrsfs}
\usepackage{enumerate}
\usepackage{amssymb}
\usepackage{algorithm}
\usepackage{algpseudocode}

\usepackage{array}
\usepackage{underscore}
\usepackage{url}

\usepackage{flushend}

\usepackage[square, comma, sort&compress, numbers]{natbib}
\usepackage{graphicx}
\usepackage{epstopdf}
\usepackage{diagbox}

\usepackage{caption}

\usepackage{subfigure}
\usepackage{color}
\definecolor{markcolor}{rgb}{0, 0, 1}
\definecolor{backcolor}{rgb}{0, 0, 0}

\newcommand{\umk}{\color{backcolor}}

\usepackage[outerbars,color]{changebar}
\ifx\pdfoutput\undefined
\else\ifnum\pdfoutput>0
\usepackage{pdfcolmk}
\fi\fi
\usepackage{fancyhdr}

\makeatletter
\newcommand{\Rmnum}[1]{\expandafter\@slowromancap\romannumeral #1@}
\makeatother

\usepackage{array}  \newcommand{\PreserveBackslash}[1]{\let\temp=\\#1\let\\=\temp}  \newcolumntype{C}[1]{>{\PreserveBackslash\centering}p{#1}}  \newcolumntype{R}[1]{>{\PreserveBackslash\raggedleft}p{#1}}  \newcolumntype{L}[1]{>{\PreserveBackslash\raggedright}p{#1}}

\ifCLASSINFOpdf
\else
\fi

\makeatletter
\def\ps@IEEEtitlepagestyle{%
	\def\@oddfoot{\mycopyrightnotice}%
	\def\@evenfoot{}%
}
\def\mycopyrightnotice{%
	{\hfill \tiny Copyright (c) 2020 IEEE. Personal use of this material is permitted. However, permission to use this material for any other purposes must be obtained from the IEEE by sending a request to pubs-permissions@ieee.org.\hfill}
}
\makeatother

\begin{document}

\title{ Resonant Beam Communications with \\Echo Interference Elimination}
\author{Mingliang Xiong, Qingwen Liu, Gang Wang, Georgios B. Giannakis, \\Sihai Zhang, Jinkang Zhu, and Chuan Huang

	\thanks{Manuscript received May 9, 2020; revised July 29, 2020; accepted August 28, 2020. Date of publication xxx, 2020; date
	of current version xxx, 2020. Part of this paper was presented at the 11th International Conference on Wireless Communications and Signal Processing (WCSP), Xi'an, China, October 23-25, 2019 \cite{WCSP2019}.  \emph{(Corresponding author: Qingwen Liu.)}}
\thanks{
	M. Xiong and Q. Liu
	are with the College of Electronics and Information Engineering, Tongji University, Shanghai 201804, China. 
	(e-mail: xiongml@tongji.edu.cn; qliu@tongji.edu.cn).}
\thanks{
	G. Wang is with the State Key Lab of Intelligent Control and Decision of Complex Systems and the School of Automation, Beijing Institute of Technology, Beijing 100081, China. (e-mail: gangwang@bit.edu.cn).}
	\thanks{G. B. Giannakis
	is with the Department of Electrical and Computer Engineering, University of Minnesota, Minneapolis, MN 55455, USA
	(e-mail: georgios@umn.edu).}
\thanks{
	S. Zhang and J. Zhu
	are with the Key Lab of Wireless-Optical Communications, University of Science and Technology of China, Hefei, Anhui 230026, China
	(e-mail: shzhang@ustc.edu.cn; jkzhu@ustc.edu.cn).}

\thanks{
	C. Huang is with the School of Science and Engineering, The Chinese University of Hong Kong, Shenzhen, Guangdong 518172, China (e-mail: huangchuan@cuhk.edu.cn).}
}

\maketitle

\begin{abstract}
	 Resonant beam communications (RBCom) is capable of providing wide bandwidth when using light as the carrier. Besides, the RBCom system possesses the characteristics of mobility, high signal-to-noise ratio (SNR), and multiplexing. Nevertheless, the channel of the RBCom system is distinct from other light communication technologies due to the echo interference issue. In this paper,  we reveal the mechanism  of the  echo interference and propose the method to eliminate the interference. Moreover, we present an exemplary design based on frequency shifting and optical filtering, along with its mathematic model and performance analysis. The numerical evaluation shows that the channel capacity is greater than $15$ bit/s/Hz.
\end{abstract}



\section{Introduction}
\label{sec:intro}

\IEEEPARstart{T}{he}
rapid advancement of mobile communications has been along with the increased carrier frequency. According to Shannon's law, wider bandwidth supports higher channel capacity. Therefore, the sixth generation (6G) mobile communications is expected to operate at terahertz~\cite{a180820.09,fdeng1,3218757,tccn20219ali}.

Wireless optical communications~(WOC) that uses infrared or visible light is a promising technology, as its carrier frequency is up to several hundred terahertz~\cite{8550716}. Light emitting diodes~(LEDs) and lasers are the most common light emitters employed in WOC~\cite{a180730.03,a180718.02}. Depending on the functionality, they can be categorized into non-line-of-sight (NLOS) and line-of-sight~(LOS) WOC~\cite{a190505.06}.  The former is based on diffuse reflection of ceilings and walls, leading to low received power and low signal-to-noise ratio~(SNR).

Most LOS WOC transmitters focus their LED radiation or laser beams on the receiver to obtain a high-level SNR. However, it is generally difficult for these directional transmitters to track the receiver. Micro-electromechanical system (MEMS)-actuated beam steering employed in many research exhibit a slow response for  tracking~\cite{a190611.01,a190505.04}.

\begin{figure}
	\centering
	\includegraphics[width=2.6in]{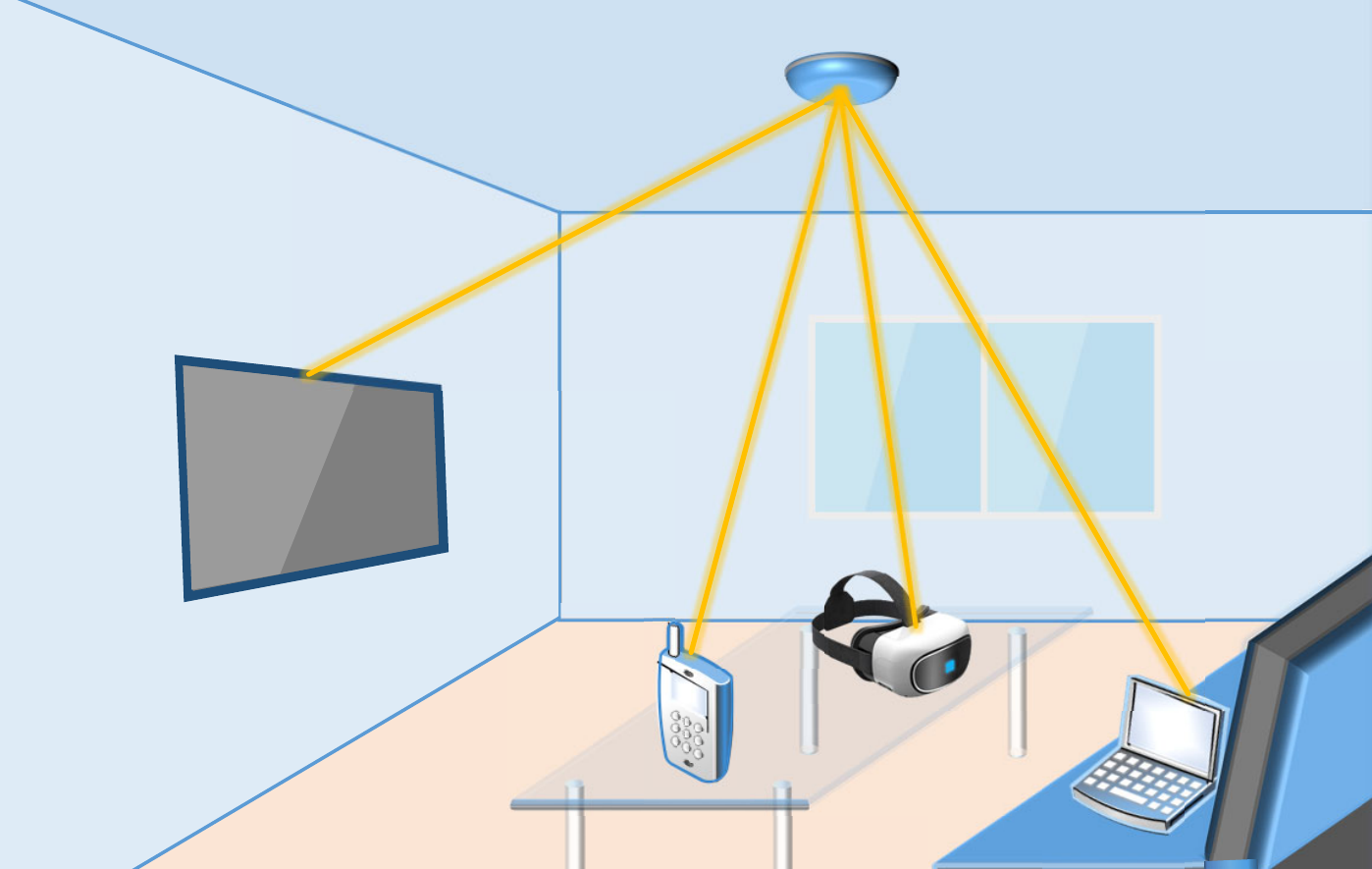}
	\caption{Scenario of resonant beam communications}
	\label{fig:Scenario}
\end{figure}

Employing non-mechanical beam steering, some optical mobile communication~(OMC) systems can response with a high speed and accuracy. For example, some research employs static diffraction gratings for fast beam steering~\cite{a190611.03}. The direction of the beam is controlled through changing the beam wavelength. Besides, a two-dimensional~(2D) multiple-beam steering is implemented with a pair of crossed gratings as well as a multiple-wavelength beam. An alternative beam steering relies on spacial light modulators~(SLMs)~\cite{a190505.01}. Each pixel of the SLM plane can be controlled to change the phase of the light that passes through this pixel. A complex light field containing multiple beams can be generated by the SLM for point-to-multipoint communication. These technologies need high-accuracy localization. However, radio frequency~(RF) for localization usually undergoes electromagnetic interference~(EMI)~\cite{79289}.

Resonant beam communications (RBCom) has the capability of high SNR, mobility, and multiplexing. The transmitter can generate resonant beams to receivers automatically, without a software or hardware controller, as depicted in Fig.~\ref{fig:Scenario}. The frequency of the resonant beam is up to $282~\mbox{THz}$.

Resonant beam system~(RBS) is the core structure of the RBCom system, and it features self-alignment, high power, and multiple beams; see~\cite{a180727.01,a180820.07,qqzhang2019,WCSP2019}. RBS is suitable for scenarios where high-power and long-distance power transmission are required, which is different from RF-based power harvesting technologies~\cite{fdeng2,HYang2020,LiuXiao2019,LiHe2018}. In RBS, a spatially separated laser resonant cavity is formed by the transmitter and the receiver~\cite{a190318.02,a190318.01}. In this cavity, the resonant beam oscillating back and forth acts as the transmission carrier.  With the retro-reflectors at both transmitter and receiver, the RBCom transmitter is capable of tracking receivers automatically. However, it also results in echo interference.

In this paper, we elaborate on the mechanism of echo interference in the RBCom system and reveal the  causes of the interference including the reflected signal, the backward modulation, and the gain fluctuation. Afterwards, we present the method of eliminating the echo interference, and propose an exemplary design as well as its mathematic channel model.

The rest of this paper is organized as follows. In Section~\ref{sec:rbcom}, we present the mechanism of RBCom. In Section~\ref{sec:echochan}, we analyze the causes of echo interference in the RBCom channel. In Section~\ref{sec:Interf}, we illustrate the echo interference elimination method based on frequency shifting and optical filtering. In Section~\ref{sec:exem}, we demonstrate the interference-free RBCom system design as well as the performance analysis. At last, we make the conclusions in Section~\ref{sec:con}.

\section{Resonant beam communications}
\label{sec:rbcom}

\begin{figure}
	\centering
	\includegraphics[width=3.4in]{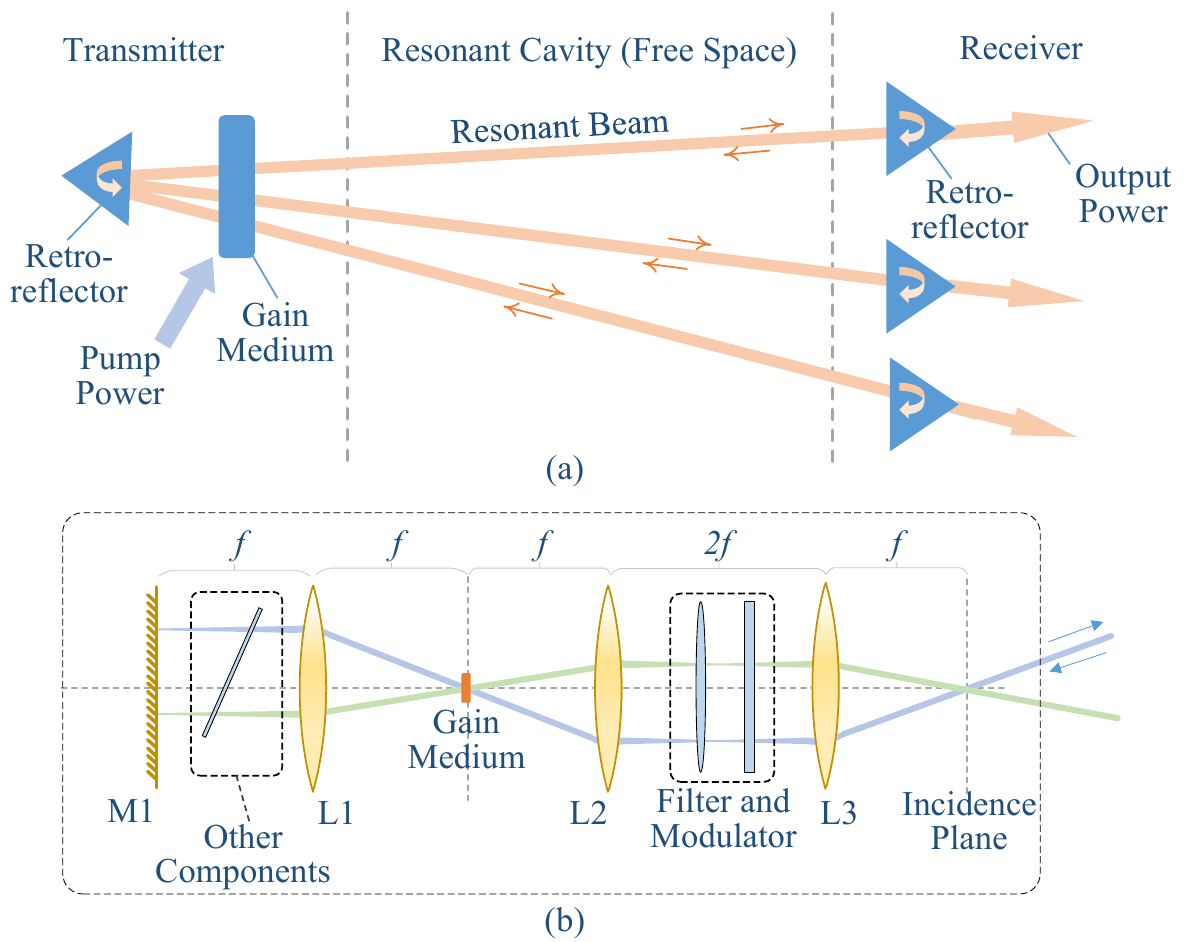}
	\caption{Resonant beam system: (a) system structure; (b) transmitter design based on cat's eye structure}
	\label{fig:rbs}
\end{figure}

Based on the RBS system, RBCom transmitter sends resonant beams to multiple receivers. By modulating the resonant beams, high-SNR mobile communications can be achieved. The RBS cavity is an open laser  cavity and we will describe the similarities as well as differences between it and traditional lasers in this section.

In a common laser cavity, photons oscillate between two aligned mirrors (a back mirror and a front mirror) and are amplified by the gain medium~\cite{a181218.01}. This optical amplification capability is supported by stimulated emission which is powered by the pump source (electricity or light). A small proportion of oscillating photons pass through the front mirror (transmittance is larger than $0$) to yield a laser beam. Note that the two mirrors of the laser cavity have to be aligned accurately, otherwise the oscillation can not be maintained.

In Fig.~\ref{fig:rbs} (a),
using two retroreflectors at both ends of the RBS, this design provides the capability of self-alignment, supporting the mobility of RBS. Retroreflectors can reflect the incident beam back along the incident direction~\cite{a180805.02,a180805.06}. Therefore, there is a path for photon oscillating between two retroreflectors. The resonant beam is formed in this oscillating path regardless of the position and the attitude of the retroreflectors. The two retroreflectors is aligned automatically, so the intra-cavity resonant beams exist even when the receivers are moving.

In addition to the RBS structure, the RBCom system has some necessary optical components including electro-optic modulator~(EOM), optical filter, and splitter. These components usually only operate at a fixed beam incidence angle, which can be satisfied by the cat's eye structure presented in~\cite{a181214.05}. Figure.~\ref{fig:rbs}(b) demonstrates a transmitter design based on the cat's eye structure. All the outside beams that pass the focal point of the lens L3 and enter the transmitter can be reflected to their source. In the body of this transmitter structure, two special zones satisfy the needs of the aforementioned optical components, i.e., between the mirror M1 and the lens L1, and between the lenses L2 and L3. Beams in these zones are parallel to the optical axis, so a fixed incident angle onto these optical components can be guaranteed. Moreover, other optical components can be integrated into the transmitter to take charge of operations including the selection of transverse and longitudinal modes.

Owing to the spatially separated resonant cavity, RBCom features safety, mobility, and multiplexing, which are elaborated as follows~\cite{a8875710}.
\begin{itemize}
	\item\emph{High-SNR and Safe}: The pencil beam generated in the resonant cavity carries high-power optical energy to the receiver, enabling high-SNR communication. When foreign objects enter the beam path, the highest-order transverse modes stop oscillating due to increased diffraction loss, and the power is transferred to lower-order modes. With the deep-entering of the foreign objects, lower-order modes will vanish. As the lowest modes stop oscillating, the resonant beam ceases, ensuring safety.
	
	\item\emph{Mobility}: Due to the self-alignment capability of retroreflector,  a robust cavity can be created for forming a resonant beam even when the transmitter and the receiver change their positions. Hence, mobile communications can be enabled.
	
	\item\emph{Multiplexing}: One transmitter can be connected with multiple receivers. For each receiver, an individual resonant cavity is created within which a resonant beam is formed. This feature, cooperating with time-division multiplexing access (TDMA) scheme or SLM array,  supports multi-user communications.
\end{itemize}

	The channel of the RBCom system is different from that of other communication systems. The carrier beam profile is determined by the resonating scheme, which means that the beam radius and the transverse intensity pattern are adjusted automatically to improve the transmission efficiency. Modulation is performed by the EOM located on the beam path. The electric field amplitude of the carrier beam is modulated during the modulation procedure. However, the retroreflector at the receiver reflects the signal back to the transmitter, affecting the modulation procedure, which is detailed in the next section.
\umk
\section{echo-interfered Channel}
\label{sec:echochan}
\begin{figure}
	\centering
	\includegraphics[width=3.1in]{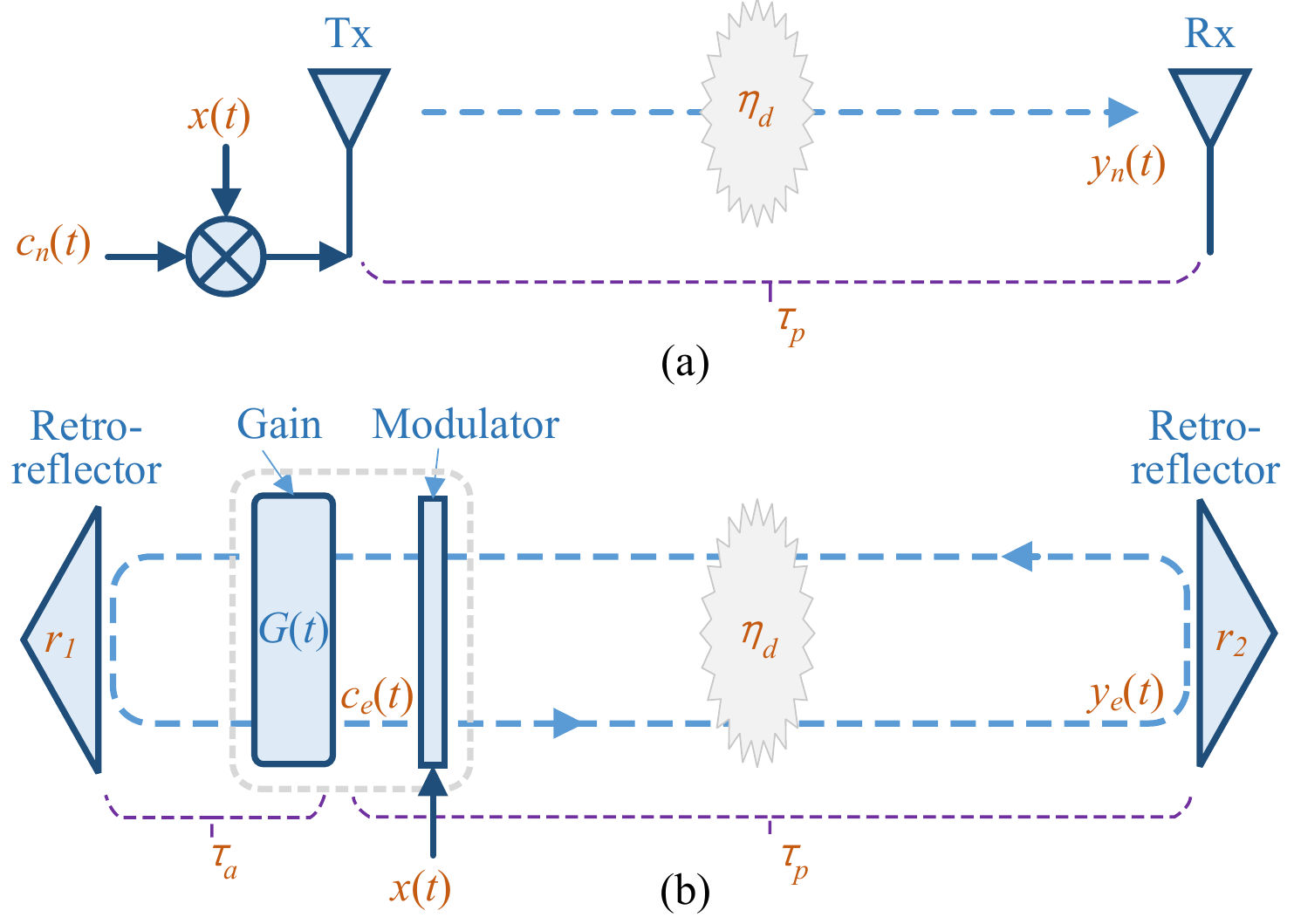}
	\caption{Channel comparison of (a) non-echo case and (b) echo-interfered case }
	\label{fig:EchoChn}
\end{figure}

In this section, the channel characteristics of RBCom are analyzed. Figure~\ref{fig:EchoChn} (a) and (b) depict a conventional non-echo communication channel and a RBCom channel, respectively. Both of them are used to transmit signals from the transmitter (the left-hand side) to the receiver (the right-hand side). Nevertheless, the RBCom channel undergoes echo interference, which makes it completely different from the non-echo channel.

In conventional communication systems, the channel does not exhibit  echo interference. As shown in Fig.~\ref{fig:EchoChn} (a), the transmit antenna~(Tx) sends the modulated signal to the receive antenna~(Rx). The carrier signal $c_n(t)$ is monochromatic with the amplitude that is a constant. The source signal is $x(t)$. The modulated signal is the product of the carrier signal and the source signal.

In contrast, the channel of RBCom systems has echo interference. As shown in Fig.~\ref{fig:EchoChn} (b), the carrier beam in the resonant cavity is expressed as
\begin{equation}
c_e(t)= A_c(t)e^{i\omega_c t}
\label{equ:rbcom-carr}
\end{equation}
where $i:=\sqrt{-1}$,  $A_c(t)$ is the amplitude, and $\omega_c$ is the beam frequency. Here we assume a single-frequency carrier beam which can be obtained by adding mode-selecting components into the cavity. The modulated signal travels circularly in the cavity. In a single round trip, a proportion of the received signal is reflected back to the transmitter, constituting the echo signal. Then, the echo signal is changed by the modulator and amplified by the gain medium, respectively. Afterwards, the transmitter's retroreflector reflects the echo to the gain medium. Finally, the echo reaches at the modulator, serving as the carrier. The echo will be modulated subsequently. Hence, the amplitude of the carrier is time-variant; that is
\begin{equation}
	A_c(t)=r_1 r_2 \eta_d y_e(t-\tau_p-2\tau_a)x(t-2\tau_a)  G(t-2\tau_a)
	G(t)
\end{equation}
where $r_1$ and $r_2$  are the retroreflectors' reflection coefficients (intensity reflectivity are $r_1^2$ and $r_2^2$); $\eta_d$ is the  transmission coefficient for the beam to travel from one side to another side of the cavity (in this paper, the term \emph{transmission coefficient} is used to describe the amplitude change ratio of a wave passing through an object such as a mass of air or a lens); $G(t)$ is the time-variant gain provided by the gain medium; $\tau_p$ is the single-pass transmission time between the transmitter's modulator and the receiver; and  $\tau_a$ is the time that signals travel from the retroreflector at the transmitter to the modulator. For simplicity, the thickness of the gain medium and the modulator is neglected. The interval between the medium and the modulator is also neglected. The cavity length is $L=(\tau_a+\tau_p)v$, where $v$ is the light speed.

\begin{figure}
	\centering
	\includegraphics[width=2.43in]{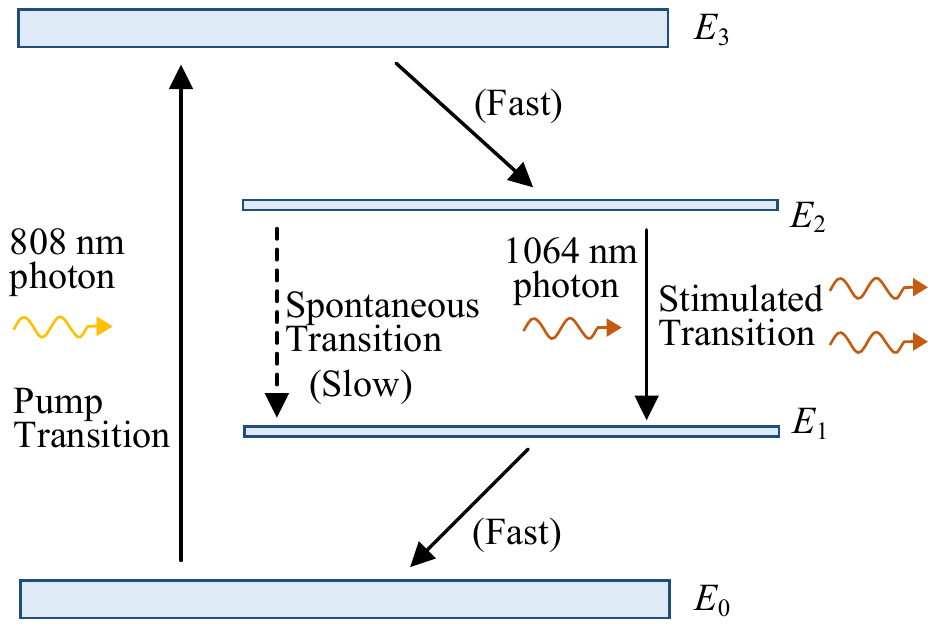}
	\caption{Simplified energy level diagram of a four-level system}
	\label{fig:energylevel}
\end{figure}

The modulated signal is the product of the carrier amplitude and the modulator's transparency. Hence, the amplitude of the received signal is
\begin{align}
y_e(t)&=\eta_dx(t-\tau_p)A_c(t-\tau_p) \nonumber\\
&=r_1 r_2\eta_d^2 y_e(t-2\tau_p-2\tau_a)x(t-\tau_p-2\tau_a) \nonumber\\
&~~~\times G(t-\tau_p) G(t-\tau_p-2\tau_a)x(t-\tau_p).
\label{equ:rbcom-recvsig}
\end{align}
The term $x(t-\tau_p)$ is the source signal to be transmitted. It is clear that the received signal is interfered by the the following aspects. The first interference is the previous received signal, which is  $y_e(t-2\tau_p-2\tau_a)$ in \eqref{equ:rbcom-recvsig}. The second interference is produced by the modulator, as the backwards echo coming from the receiver is also changed by the modulator. Thus, the echo is multiplied by $x(t-\tau_p-2\tau_a)$. The last impact stems from the gain medium due to its time-varying gain. The echo is amplified twice through the gain medium. Hence, the echo is multiplied by  $G(t-\tau_p)G(t-\tau_p-2\tau_a)$.

The RBCom cavity is a time-variant system. So gain variation can be observed if there is any influence imposed on the intra-cavity resonant beam. As depicted in Fig.~\ref{fig:energylevel}, the gain medium is often modeled as a four-level system with four energy levels $\{E_0,E_1,E_2,E_3\}$. As the pump light (e.g., $808$ nm) enters the gain medium, the populations at $E_0$ transit to $E_3$ through stimulated absorption, and then quickly transit to $E_2$. As the oscillating photons (e.g. $1064$ nm) enter the gain medium, the populations at $E_2$ transit to $E_1$, generating extra $1064$ nm photons through stimulated emission.  Generally, the population densities at $E_1$ and  $E_3$ are assumed to be $0$, as the lifetime of the populations at energy level $E_1$ and $E_3$ is much shorter than that of populations at other energy levels~\cite{a181218.01}. Let $n_0$ ($n_2$) denote the density of populations at $E_0$ ($E_2$), the aforementioned phenomenon is described mathematically by the following simplified rate equations~\cite{a181218.01,a181228.01}
\begin{align}
\label{equ:rate-equ}	\displaystyle\frac{\partial n_2}{\partial t}&= - n_2\phi\sigma v -\frac{n_2}{\tau_f} + R_p \\
\label{equ:rate-equ2}	\displaystyle\frac{\partial \phi}{\partial t}&=  n_2\phi\sigma v -\frac{\phi}{\tau_c} + S
\end{align}
where $\phi$ is the photon density; $\sigma$ is the stimulated emission cross section; $\tau_f$ is the fluorescence lifetime; $R_p$ is the pump rate depending on the pump power; $\tau_c$ is the photons lifetime; and $S$ is the spontaneous emission rate for  photons to be coupled into the cavity. The equations \eqref{equ:rate-equ} and \eqref{equ:rate-equ2} describe the change rate of $n_2$ and $\phi$, respectively.

Stimulated emission results in a decrease of $n_2$ and an increase of $\phi$~\cite{a181218.01}. Simultaneously, the stimulated emission rate is also determined by $n_2$ and $\phi$, as $ n_2\phi\sigma v$ in  \eqref{equ:rate-equ} and \eqref{equ:rate-equ2} represents the stimulated emission rate. During this process, the populations at energy level $E_2$ decay to the ground level $E_0$ through generating an equivalent number of photons. The photons are consumed gradually, as there are air absorption and diffraction loss in the cavity. The density $n_2$ of populations at energy level $E_2$ is supplied by the stimulated absorption transition of the populations at $E_0$ with rate $R_p$, which is related to  $n_0$ and the pump power $P_{\rm in}$. Generally, if $P_{\rm in}$ is stable, $R_p$ can be assumed as a constant, since $n_0\gg n_2$. A negative feedback mechanism can be observed in the rate equation system. Given adequate $P_{\rm in}$, the rate equation system can reach the steady state, in which $\partial{\phi}/\partial{t}=\partial{n_2}/\partial{t}=0$~\cite{a181218.01}. A quasi-monochromatic resonant beam can be obtained when the system becomes stable. For simplicity, we assume the resonant beam at the steady state is a monochromatic wave.

\begin{figure}
	\centering
	\includegraphics[width=3.2in]{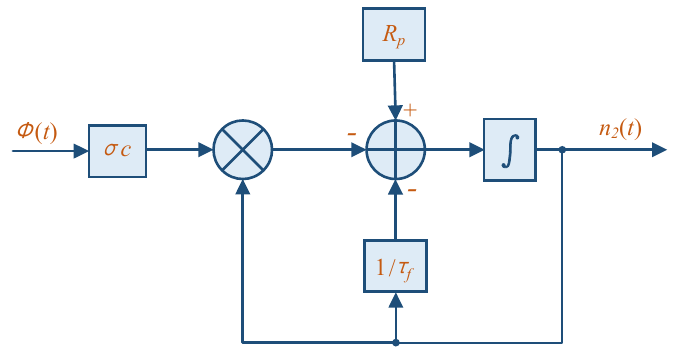}
	\caption{Upper-level population density interacting with photon density}
	\label{fig:Gain}
\end{figure}

The upper-level population density $n_2$ determines the amplitude gain provided by the gain medium. This relationship is expressed as  ~\cite{a181218.01}
\begin{equation}
G(t)= \sqrt{e^{ n_2(t) \sigma l}}
\end{equation}
where $l$ is the gain medium length. Figure~\ref{fig:Gain} depicts how $\phi$ affects $n_2$. Moreover, both $n_2(t)$ and $G(t)$ are non-linear systems related to the pump rate $R_p$. Consequently, gain fluctuation exists when the pump source of the gain medium has a time-variant power.

In a nutshell, the RBCom channel is difficult to be utilized for communications directly. The following three issues need to be addressed for establishing a non-interference resonant beam channel.
\begin{itemize}
	\item\emph{Reflected Signal}: The echo beam reflected by the receiver is comprised of information components, affecting the subsequent modulation. The echo should not be cut off arbitrarily as it plays a key role in maintaining the resonance. An effective way to eliminate the information signal in the echo is to remove its components in the frequency domain.
	
	\item\emph{Backward Modulation}: Not only the forward beam is changed by the modulator, the reflected echo beam is also changed as it passes through the modulator. Even if the echo is pure with only carrier frequency, other frequency will be added to the echo when it passes through the modulator. The frequency components induced by the backward modulation need to be removed.
	
	\item\emph{Gain Fluctuation}: Only under a balance condition can the gain become stable, where the upper-level population density as well as the photon density is no longer changed. Any impacts applied to the gain medium will results in gain fluctuation. Hence, the pump power as well as the  stimulated photon density in the cavity should  be a static vaule during the communication process.
\end{itemize}
	
\section{Interference Elimination}
\label{sec:Interf}

\begin{figure}
	\centering
	\includegraphics[width=3.0in]{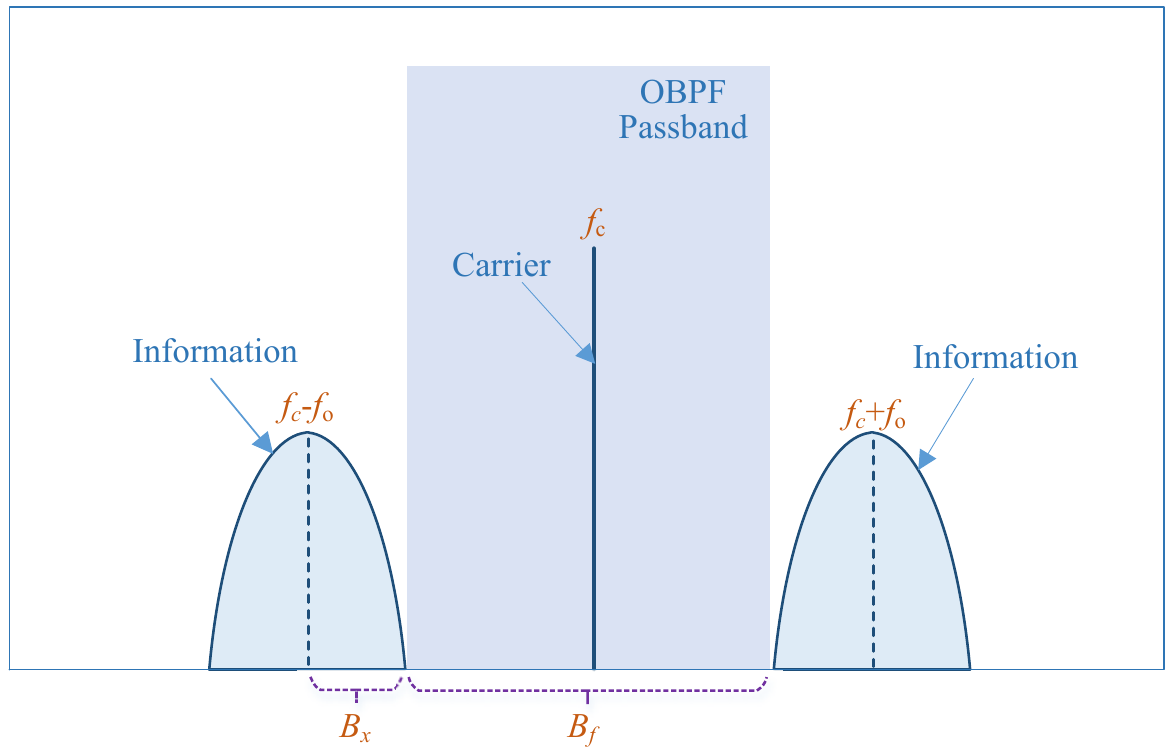}
	\caption{Interference elimination with optical filtering}
	\label{fig:FreqShift}
\end{figure}

The RBCom channel undergoes echo interference, and therefore can not be utilized directly for data transfer. Yet, it is able to eliminate the echo and create a well-performed non-interference channel. For this purpose, the information components in the echo should be removed; the impacts of the backward modulation should be avoided; and the gain stability should be maintained. Two procedures presented below is adopted to meet these requirements.

\subsection{Optical Filtering}

 As demonstrated in Fig.~\ref{fig:FreqShift}, the desired spectrum pattern of the modulated signal should allow the information bands to be completely filtered out, which is described as
\begin{align}
Y(f)&=p' C_0(f)+\dfrac{m'}{4}X(f+f_c+f_o)+\dfrac{m'}{4}X(f+f_c-f_o) \nonumber\\
&~~~+\dfrac{m'}{4}X(f-f_c+f_o)+\dfrac{m'}{4}X(f-f_c-f_o).
\label{equ:desire-Y}
\end{align}
The modulated signal $Y(f)$ is comprised of the information bands $X(f)$ and the carrier components $C_0(f)$. The carrier is a monochromatic wave with frequency of $f_c$ as the resonance reaches stable. Thus, the carrier's frequency spectrum is
\begin{equation}
C_0(f)=\dfrac{1}{2}\delta(f-f_c)+\dfrac{1}{2}\delta(f+f_c)
\end{equation}
where $\delta(f)$ denotes the unit-impulse function.
As the positive and the negative frequency axis are symmetrical, Fig.~\ref{fig:FreqShift} only demonstrates the positive  axis. Two frequency bands carrying information sit symmetrically over $f_c$, at frequencies $f_c\pm f_o$, respectively.

The OBPF has a central frequency $f_c$, and a bandwidth $B_f$. Hence, the frequency response of the OBPF is
\begin{equation}
H_{\rm OBPF}(f)=\Pi\left(\frac{f+f_c}{B_{ f}}\right)+\Pi\left(\frac{f-f_c}{B_{f}}\right)
\end{equation}
where $\Pi(f)$ is the rectangular function  described as
\begin{equation}
\Pi(f)=\left\{
\begin{array}{l l}
1, & |f| \le 1/2\\
0, & |f| > 1/2\\
\end{array}
\right..
\end{equation}
The information bands  can be removed only when they are out of the filter's passband. Hence, the following condition needs to be met; that is
\begin{equation}
0<B_{ f}<2f_o-2B_{ x}
\end{equation}
where $B_x$ is the source signal bandwidth. After filtering, only carrier frequency exists in the echo, namely
\begin{equation}
Y'(f)=Y(f)H_{\rm OBPF}(f)=p' C_0(f).
\label{equ:Y-filtered}
\end{equation}
According to \eqref{equ:Y-filtered}, OBPFs detach the information components from the echo. The carrier frequency in the echo is preserved for maintaining the beam resonance. In this case, the previous transmitted information signal can not interfere the ongoing modulation. In the same way, the impact of the backward modulation can be eliminated.

\subsection{Maintaining Gain Stability}

The power loss of the photons in the cavity is compensated by the gain medium. In order to generate a resonant beam, the gain of the gain medium must be greater than the total loss in the cavity at the beginning of the communication progress. On this premise, the photon density in the cavity increases gradually, which also speeds up the consumption of the populations at energy level $E_2$ in the gain medium. In the meantime, the gain decreases gradually until becoming equal to the loss. At last, the system achieves a steady state where the gain as well as the upper-level population density $n_2$ will not change. This process is based on the negative feedback mechanism of the gain system.

During the modulation, the gain system has to be kept in the steady state, for which two requirements introduced in the following should be satisfied.
\begin{itemize}
	\item \emph{Steady Photon Density in Gain Medium}: In Fig.~\ref{fig:Gain}, the steady state of the gain system will be broken when meeting photon density fluctuation. In this case, the gain is unexpectedly dynamic. Consequently, the echo toward the gain medium is expected to have a constant intensity~(i.e., a constant photon density). As depicted in~\eqref{equ:Y-filtered}, a monochromatic wave that has constant amplitude and constant intensity is obtained.
	
	\item \emph{Threshold Condition}: During each round trip, the gain is required to be greater than the loss; otherwise, the resonance cannot be built up~\cite{a181218.01}. Hence, the round-trip transmission coefficient $\mathcal{G}_{r}$  must satisfy
	\begin{equation}
	\mathcal{G}_{r}(t)=\dfrac{c_e'(t)}{c_e'(t-2\tau_a-2\tau_p)}=r_1 r_2 \eta_m^2 \eta_d^2  G^2(t) \geq 1
	\label{equ:loopG}
	\end{equation}
	where $\eta_m$ is the transmission coefficient of the modulator which is determined by the operating point; and $c_e'(t)$ is the carrier processed by OBPFs. We assume the OBPF is an ideal device, so the attenuation in the passband is neglected. The steady state where $\mathcal{G}_r(t)=1$ is obtained automatically because of the negative feedback mechanism mentioned above. Then, the steady-state gain is given as follows
	\begin{equation}
	G=G(t)\Big|_{\mathcal{G}_r(t)=1}=\dfrac{1}{\eta_m \eta_d\sqrt{r_1 r_2}}.
	\label{equ:staticG}
	\end{equation}

\end{itemize}

\section{Exemplary Design}
\label{sec:exem}

In the above section, two procedures are proposed for interference elimination, i.e., optical filtering and gain stability maintenance. These interference-eliminating procedures are demonstrated next with an exemplary  system design.

As shown in Fig.~\ref{fig:EchoFreeDesn}, before entering the electro-optic amplitude modulator, the source signal $mx(t)$ (the  amplitude $m\ge0$ and $|x(t)|\le1$) is preprocessed for meeting the desired spectrum pattern. At the receiver, a splitter is used to reflect a fraction of the resonant beam to the photon detector~(PD) for demodulation, and the rest of the beam that passes through the splitter is reflected back to the transmitter by a retroreflector for maintaining the resonance. The OBPF behind the splitter is adopted to detach the information bands from the echo's spectrum. Between the gain medium and the modulator, another OBPF is mounted to detach the undesired frequency generated through backward modulation. Modulation,
demodulation, and gain stability maintenance are detailed below.

\subsection{Modulation}

\begin{figure*}
	\centering
	\includegraphics[width=5.0in]{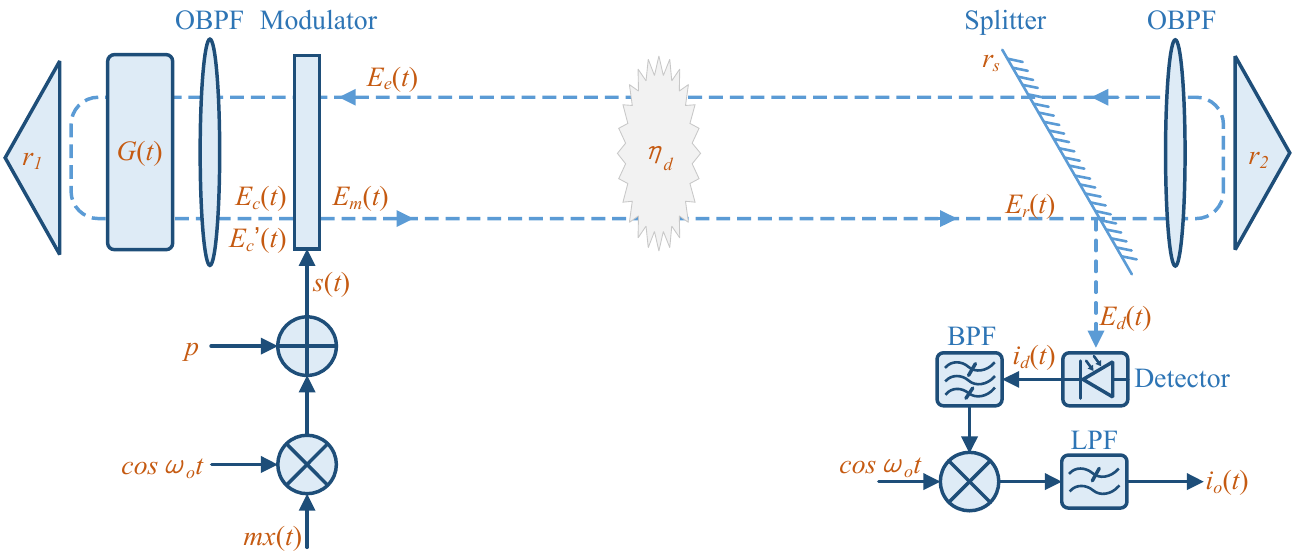}
	\caption{Exemplary resonant beam communication system design with echo elimination}
	\label{fig:EchoFreeDesn}
\end{figure*}

In order to ensure that the source signal band is away from zero frequency, the  band of the source signal should be shifted by $\omega_o/2\pi$. According to Fig.~\ref{fig:EchoFreeDesn},  the local oscillation $\cos\omega_o t$ and the source signal $mx(t)$ is mixed up by a mixer to conduct frequency shifting. Next, the operating point for the modulator is specified by adding a direct-current (DC) bias $p$ to the processed signal.

When the RBCom achieves the steady state where the carrier amplitude $A_c$ is a constant, the modulation is able to be initiated. Then, the field of the carrier beam is depicted as
\begin{equation}
 E_c(t)=\mbox{Re}\left[A_ce^{i\omega_c t}\right]
\end{equation}
where $\omega_c$ represents the angular frequency, and  $\mbox{Re}[\cdot]$ takes the real part. Hence, the  modulated beam's field is expressed as
\begin{equation}
E_{m}(t)=\left[mx(t)\cos{\omega_o t}+p\right] E_c(t)
\end{equation}
where $0<p-m$ and $p+m<1$; and both $p$ and $m$ are defined as positive values. To give full play to the modulator's performance, we choose $p+m=1$, thus, the modulation depth $m$ is constrained by operating point $p$.

The modulated beam experiences amplitude attenuation with ratio $\eta_d=\eta_{\rm diff}\exp(-\alpha d/2)$ as transfered from the transmitter to the receiver, where $\eta_{\rm diff}$ is the amplitude attenuation ratio induced by diffraction loss, $\alpha$ is the loss coefficient of atmosphere (for different air conditions including clear air, haze, and fog, $\alpha$ are $0.0001$~m$^{-1}$, $0.001$~m$^{-1}$, and  $0.01$~m$^{-1}$, respectively)~\cite{a200427.04}, and $d$ is the transmission distance. Hence, the received electric field at the receiver's splitter is
\begin{equation}
E_{r}(t)=\eta_d E_m(t).
\end{equation}

\subsection{Demodulation}

At the receiver, a fraction of the received beam is reflected to the PD by the splitter. Here $r_s$ denotes the splitter's reflection coefficient. The electric field at the PD is
\begin{align}
E_d(t)&=r_{s}E_{r}(t) \nonumber\\
&=\mbox{Re}\left[\tilde{E}_d(t)\right]
\end{align}
where
\begin{equation}
\tilde{E}_d(t)=r_s\eta_d A_c \left[mx(t)\cos{\omega_o t}+p\right]e^{i\omega_c t}.
\end{equation}

The PD converts the incident beam to photocurrent. According to~\cite{OptiElec5}, the generated photocurrent is proportional to the intensity of the incident light
\begin{equation}
i_d(t)\propto \tilde{E}_d(t)\tilde{E}_d^*(t)
\end{equation}
where $\tilde{E}_d^*(t)$ is the complex conjugate of $\tilde{E}_d(t)$. Consequently, the  current signal generated by the PD is
\begin{align}
i_d(t)&= k\left\{r_s \eta_d A_c \left[mx(t)\cos{\omega_o t}+p\right]\right\}^2 + n(t) \nonumber\\
&=\dfrac{1}{2} k r_s^2 \eta_d^2 A_c^2 m^2x^2(t)\cos{2\omega_o t} \nonumber\\
&~~~+ 2k r_s^2 \eta_d^2 A_c^2  pmx(t)\cos{\omega_o t} \nonumber\\
&~~~+ \dfrac{1}{2}k r_s^2 \eta_d^2 A_c^2  m^2x^2(t) \nonumber\\
&~~~+k r_s^2 \eta_d^2 A_c^2  p^2 + n(t)
\label{equ:id}
\end{align}
where $k=\eta_{\rm det}\rho \mathcal{A}_b/Z_0$ is a constant; $\rho$ is PD's responsivity; $\mathcal{A}_b$ is the cross section area of the beam; $Z_0\approx 377~\Omega$ is the impedance of free space; $n(t)$ is the PD's additive Gaussian white noise~(AWGN); and $\eta_{\rm det}$ is the ratio of the detected power to the total incident beam power. The average photocurrent $I_{\rm sig}$ contributes to the shot noise of the PD. From \eqref{equ:id}, we can obtain
\begin{equation}
	I_{\rm sig}=k r_s^2 \eta_d^2 A_c^2  p^2 + \dfrac{1}{2}k r_s^2 \eta_d^2 A_c^2  m^2 \left<x^2(t)\right>
\end{equation}
where $\left< \cdot \right>$ is the expectation operation.

The frequency band of the first three terms on the right side of \eqref{equ:id} are $[2f_o-2B_x,2f_o+2B_x]$, $[f_o-B_x,f_o+B_x]$, and $[-2B_x, 2B_x]$, respectively, where $B_x$ is the bandwidth of $x(t)$. The local oscillation frequency is $f_o=\omega_o/2\pi$. Since the source signal $x(t)$ is contained in the second term, a bandpass filter (BPF) with bandwidth of $2B_x$ is employed to extract the second term. For this purpose, the frequency band of the second term should not overlap with the frequency bands of others. Therefore, the following restriction  has to be satisfied:
\begin{equation}
\left\{
	\begin{array}{l}
		f_o+B_x<2f_o-2B_x \\
		f_o-B_x>2B_x \\
	\end{array}
\right.
\Rightarrow f_o>3B_x.
\end{equation}

Next, coherent demodulation scheme is adopted. The extracted term is mixed with $\cos{\omega_o t}$ and then processed by a lowpass filter (LPF) whose bandwidth is $B_x$ to obtained the low-frequency band.   Thus, the demodulated signal is
\begin{align}
i_o(t)=k r_s^2 \eta_d^2 A_c^2  pmx(t) + \frac{1}{2}n_b(t).
\label{equ:demod-sig}
\end{align}
where $n_b(t)$ is the band-limited white noise.

\subsection{Optical Filtering}

As demonstrated in Fig.~\ref{fig:EchoFreeDesn}, the beam passes through the OBPF twice at the receiver. The received electric field at the splitter is
\begin{align}
	E_r(t)&=\dfrac{1}{2}\eta_d A_c mx(t)\cos{(\omega_c-\omega_o)t} \nonumber\\
	&~~~+\dfrac{1}{2}\eta_d  A_c mx(t)\cos{(\omega_c+\omega_o)t} \nonumber\\
	&~~~+ p \eta_d  A_c \cos{\omega_c t}.
\end{align}
The spectrum of $E_r(t)$ meets the pattern depicted by \eqref{equ:desire-Y}. The echo field that goes into the modulator is obtained as
\begin{align}
E_e(t)&=r_2 t_s \eta_d  E_{r}(t)* h_{\rm OBPF}(t) * h_{\rm OBPF}(t) \nonumber\\
&=r_2 t_s p \eta_d^2  A_c \cos{\omega_c t}
\label{equ:em}
\end{align}
where $t_s = \sqrt{1-r_s^2}$ is the splitter's transmission coefficient, and $h_{\rm OBPF}(t)$ is the OBPF's impulse response. As depicted by \eqref{equ:em}, a monochromatic beam is obtained.

The light wave arrives at the modulator after one round trip and can be represented as
\begin{align}
E_c'(t)&=E_e(t)  s(t) * h_{\rm OBPF}(t) G^2 * h_{\rm OBPF}(t) \nonumber\\
&= r_1 r_2 t_s p^2 \eta_d^2  G^2  A_c \cos{\omega_c t}
\label{equ:new-carrier}
\end{align}
where $G$ is the cavity gain at the steady state defined in \eqref{equ:staticG}. From~\eqref{equ:new-carrier},  $E_c'(t)$ is independent of $x(t)$, which means the interference induced by the reflected signal is eliminated.

\subsection{Gain Stability}
The gain stability is not affected by modulation when the intensity of the processed echo is static. We can obtain the following round-trip transmission coefficient
\begin{align}
\mathcal{G}_{r}(t)&=\frac{E_c'(t)}{E_c(t)} \nonumber \\
&= r_1 r_2 t_s  p^2 \eta_d^2 G^2.
\label{equ:Gl-exam}
\end{align}
When $\mathcal{G}_{r}(t)=1$, the steady state is achieved. In the steady state, no matter what the source signal is, and whether the communication process starts or not, the gain of the gain medium is a constant value, namely
\begin{equation}
\displaystyle	 G=\dfrac{1}{p\eta_d \sqrt{r_1 r_2 t_s}}.
\label{equ:staticG-exam}
\end{equation}
Here, the DC bias $p$ determines the modulator's transmission coefficient $\eta_m$ in \eqref{equ:loopG}. According to \eqref{equ:staticG-exam}, $p$ affects the steady-state gain $G$.

\subsection{Numerical Examples}

In this subsection, the performance of the interference-free RBCom system is studied. The capacity $C$ and the carrier amplitude $A_{c}$ are evaluated. Since coherent demodulation is adopted, the channel  capacity can be obtained as
\begin{equation}
	C=\log_2(1+SNR)
	\label{equ:Capa}
\end{equation}
where
\begin{equation}
	SNR=\dfrac{\left< i_s^2(t) \right> }{(\sigma^2_{ shot}+\sigma^2_{ \rm thermal})/4}
\end{equation}
where $\sigma^2_{\rm shot}$ and $\sigma^2_{\rm thermal}$ are the power of the shot noise and the thermal noise of PD, respectively. From \eqref{equ:demod-sig}, the average signal power expressed in A$^2$ is obtained by
\begin{equation}
\left< i_s^2(t) \right>=k^2 r_s^4 \eta_d^4 A_c^4  p^2 m^2 \left<x^2(t)\right>.
\end{equation}
The bandwidth of the PD is $2B_x$ which is equal to the bandpass filter at the demodulation circuit. The shot noise power expressed in A$^2$ is given by
\begin{equation}
	\sigma^2_{\rm  shot}=2q(I_{\rm sig}+I_{\rm bk})\times 2B_x
\end{equation}
where $q$ is electron charge, $I_{\rm sig}$ and $I_{\rm bk}$ are the photocurrents
induced by the signal and the background radiations, respectively.
The thermal noise power expressed in A$^2$ is given by
\begin{equation}
\sigma^2_{\rm thermal}=\frac{4KT}{R_{L}}\times 2B_x
\end{equation}
where $T$ is the temperature in Kelvin and $K$ is the
Boltzmann constant, $R_{L}=10$~k$\Omega$ is the load resistor~\cite{a100424.01}.

The PD's responsivity is $\rho=0.6$~A/W and its area is $1$~cm$^2$~\cite{a200426.01}. As the beam radius is at the level of millimeter,  we assume all the beam energy is focused onto the PD by lens, i.e., $\eta_{\rm det}=1$. We set $r_1=r_2=1$, $T=298$~K. Since well coated optical components generally have ultra-low energy absorption, and typical air transmission loss is $0.01$ per $100$~m, the corresponding loss can be neglected in indoor scenario~\cite{a200427.04}. The transmission loss coefficient $\eta_d$ is dominated by diffraction loss factor which is generally greater than $0.9$ for multiple-transverse-mode operation~\cite{a181221.01}.  Hence, we set the transmission coefficient $\eta_d\approx\eta_{\rm diff}=0.949$. We assume that  $\left<x^2(t)\right>=0.3$. The background radiation induced photocurrent $I_{\rm bk}=5100~\mu$A~\cite{a200427.02}. Multiple quantum wells (MQW) modulator can support rates up to tens of gigahertz~\cite{a180805.05}. Hence, we set  $B_x=5~\mbox{GHz}$.


One can assume that the pump power is a fixed value, namely, $P_{\rm in}=30$ W. In this case, the carrier amplitude $A_c$ is then treated as a variable given by (see Appendix \ref{apdx:carrieramp} for details)
\begin{equation}
A_c=\sqrt{\dfrac{Z_0 I_s\left[\dfrac{\eta P_{\rm in}}{I_s\mathcal{A}_b}+\ln{\sqrt{t_s^2 p^4\eta_d^4}}\right]}{1-t_s^2 p^4\eta_d^4}}
\end{equation}
where $I_s$ is the saturation intensity; $\eta$ is a combined efficiency in pumping procedure; and $\mathcal{A}_b$ is the cross section area of the gain medium (We set $\mathcal{A}_b=7.854\times10^{-7}~\mbox{m}^2$, i.e., the diameter is $1~ \mbox{mm}$). The saturation intensity $I_s$ is related to the material of the gain medium. Here, we assume neodymium
doped yttrium aluminum garnet~(Nd:YAG) is employed as the gain medium. The medium's parameters are $I_s=2.901\times10^7~ \mbox{W/m}^2$,  $\sigma=2.8\times10^{-23}~\mbox{m}^2$, and  $\tau_f=230~{\upmu}\mbox{s}$. When a Nd:YAG crystal is employed as the gain medium, the wavelength of the pump light is required to be $808~\mbox{nm}$~(i.e., the pump frequency $f_p\approx371~\mbox{THz}$), and the wavelength of the stimulated beam is around $1064~\mbox{nm}$~(i.e., the carrier frequency $f_c\approx282~\mbox{THz}$). We set $\eta=0.65$, as the overlap efficiency, the quantum efficiency, and the Stocks factor are  $0.9$, $0.95$, and $0.76$, respectively~\cite{a181218.01}.

\begin{figure}
	\centering
	\includegraphics[width=3.3in]{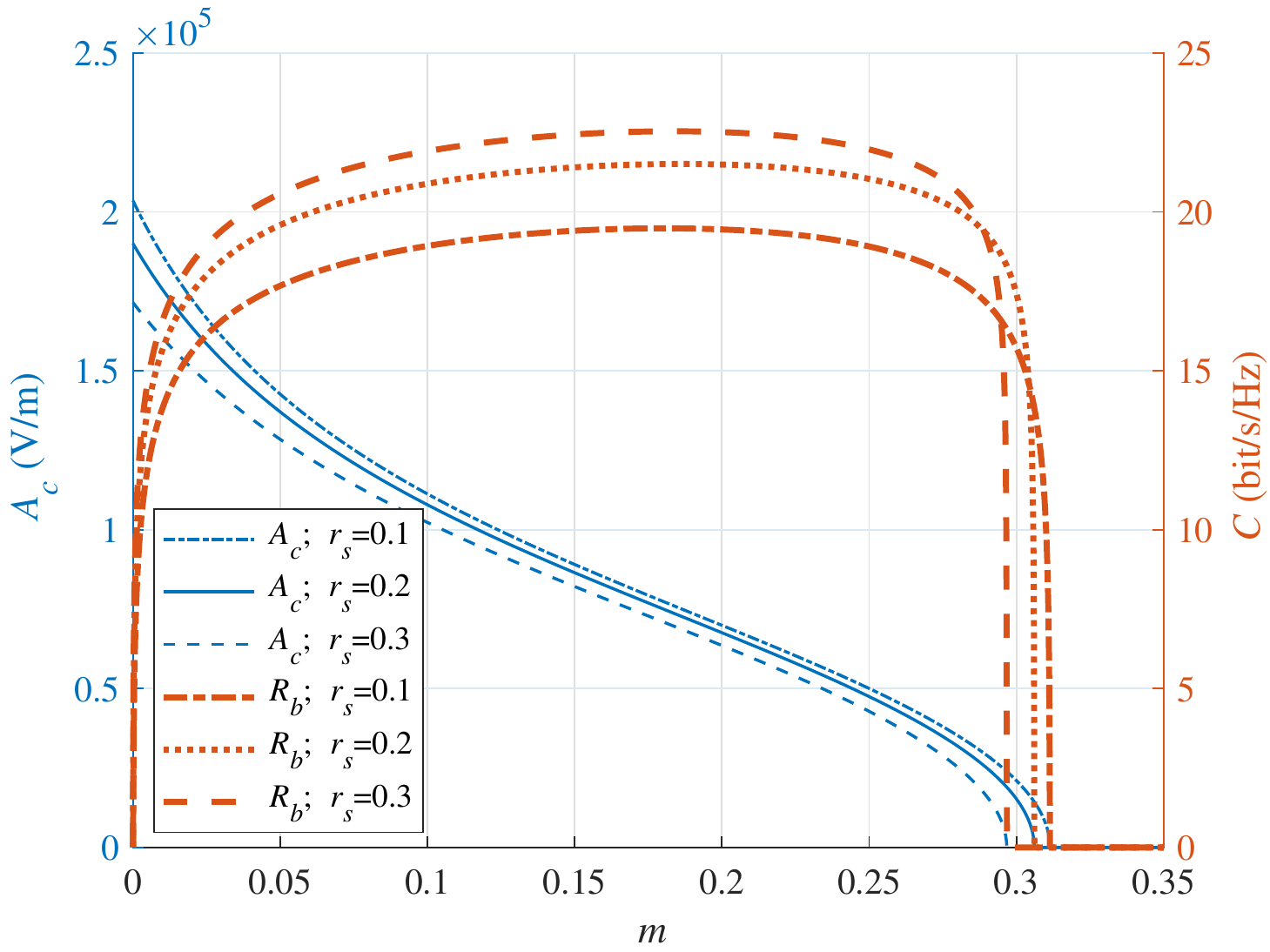}
	\caption{Carrier amplitude $A_c$ and capacity $C$ vs. peak amplitude $m$ of source signal }
	\label{fig:CurvAcRbvsm}
\end{figure}

Figure \ref{fig:CurvAcRbvsm} plots how the carrier amplitude $A_c$ (left $y$-axis) as well as the capacity $C$ (right $y$-axis) varies with the peak amplitude of the source signal $m$. As $m$ increases (i.e., the DC bias $p$ decreases), $A_c$  decreases gradually. When $m$ approaches $0$ or the threshold $m_{\rm th}$, the capacity $C$ decrease to to $0$ rapidly. A relatively high capacity  can be achieved when $m$ varies in the range between $0$ and $m_{\rm th}$. Furthermore, a smaller reflection coefficient $r_s$ of the splitter supports a greater $m_{\rm th}$.

\begin{figure}
	\centering
	\includegraphics[width=3.3in]{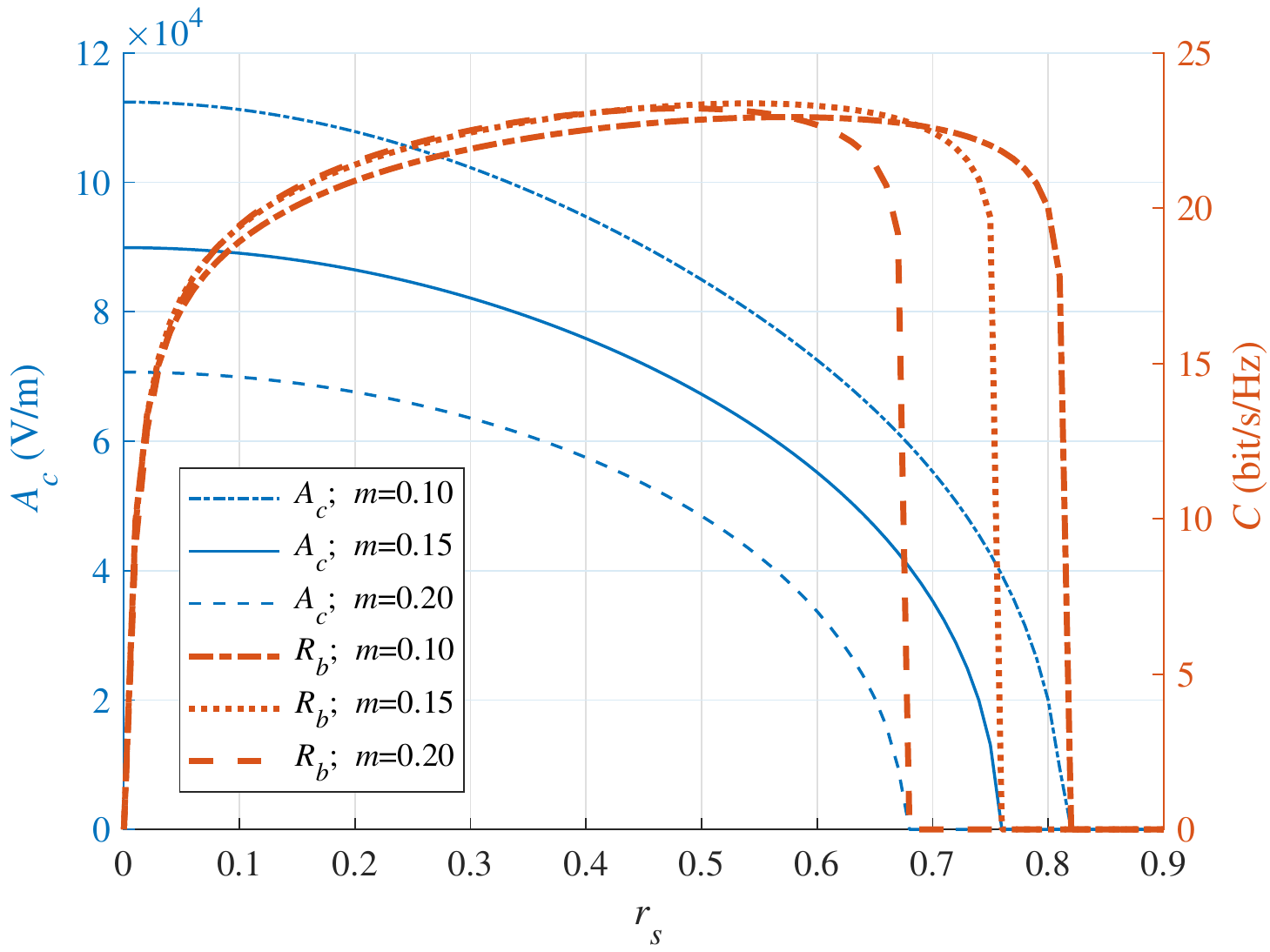}
	\caption{Carrier amplitude $A_c$ and capacity $C$ vs. splitter's reflection coefficient $r_s$}
	\label{fig:CurvAcRbvsrs}
\end{figure}

Figure \ref{fig:CurvAcRbvsrs} shows the carrier amplitude $A_c$ as a function of the reflection coefficient $r_s$ of the splitter (left $y$-axis); and also the capacity $C$ as a function of $r_s$ (right $y$-axis). As $r_s$ increases, $A_c$ decreases gradually, while $C$ raises. However, when $r_s$ approaches the threshold $r_{s,\rm th}$, $A_c$ as well as $C$ decreases to $0$ rapidly, because the total loss in the cavity becomes greater than the gain. Clearly from Fig.~\ref{fig:CurvAcRbvsrs}, a smaller $m$ (i.e., a greater DC bias $p$) leads to a greater $r_{s,\rm th}$.

\begin{figure}
	\centering
	\includegraphics[width=3.3in]{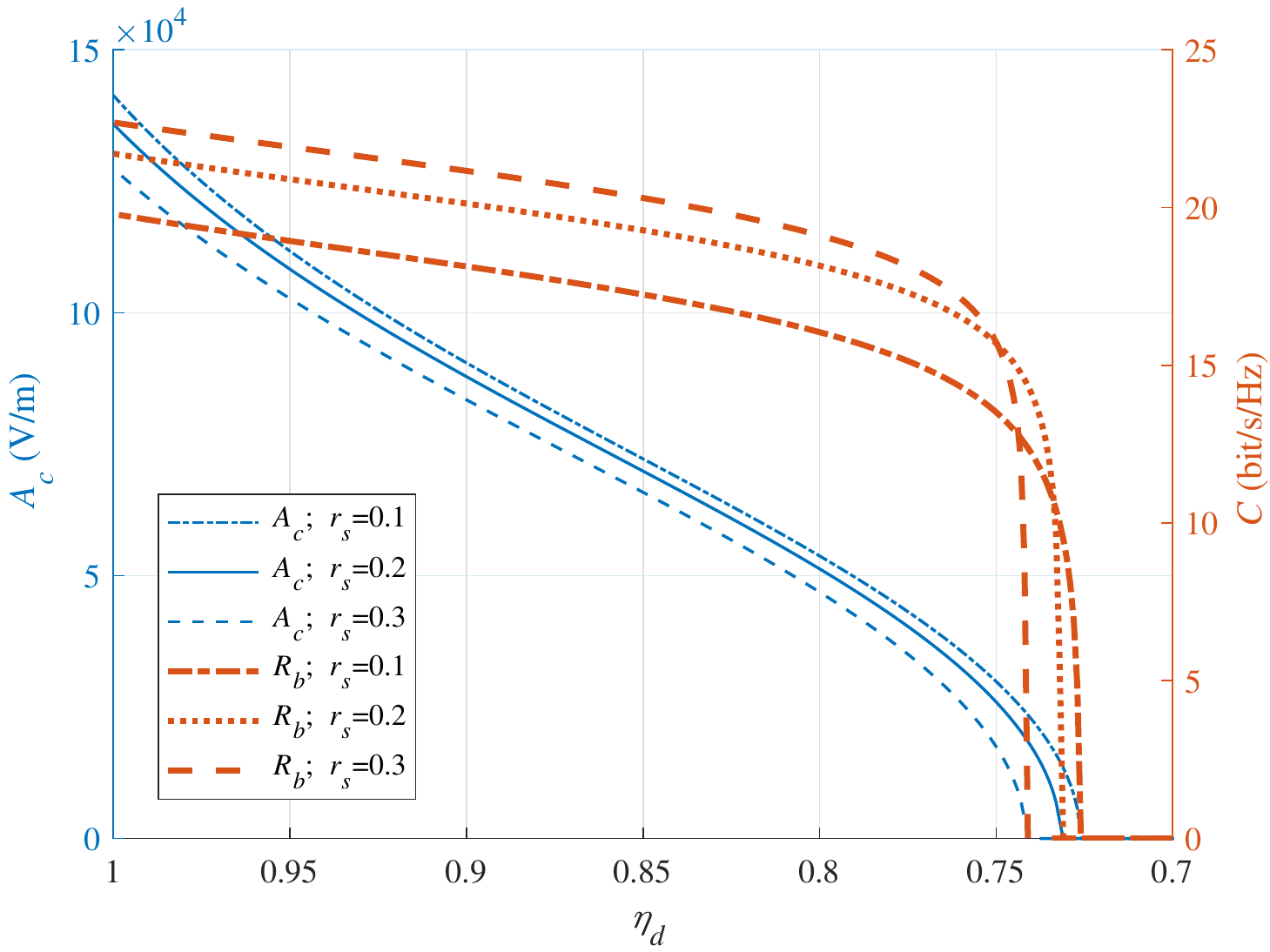}
	\caption{Carrier amplitude $A_c$ and capacity $C$ vs. spatial transmission coefficient $\eta_d$ (peak source signal amplitude $m=0.1$)}
	\label{fig:CurvAcRbvsetad}
\end{figure}

As demonstrated in Fig.~\ref{fig:CurvAcRbvsetad}, the spatial transmission coefficient $\eta_d$ also affects the carrier amplitude $A_c$  (left $y$-axis) and the capacity $C$ (right $y$-axis). Here, $m$ is set to $0.1$. With the decrease of $\eta_d$, both $A_c$ and $C$ decrease gradually. The decrease rate of $C$ is small when $\eta_d$ is great.  Nevertheless, when $\eta_d$ approaches the threshold $\eta_{d,\rm th}$, $A_c$ as well as $C$ decreases to $0$ rapidly.

\begin{figure}
	\centering
	\includegraphics[width=3.2in]{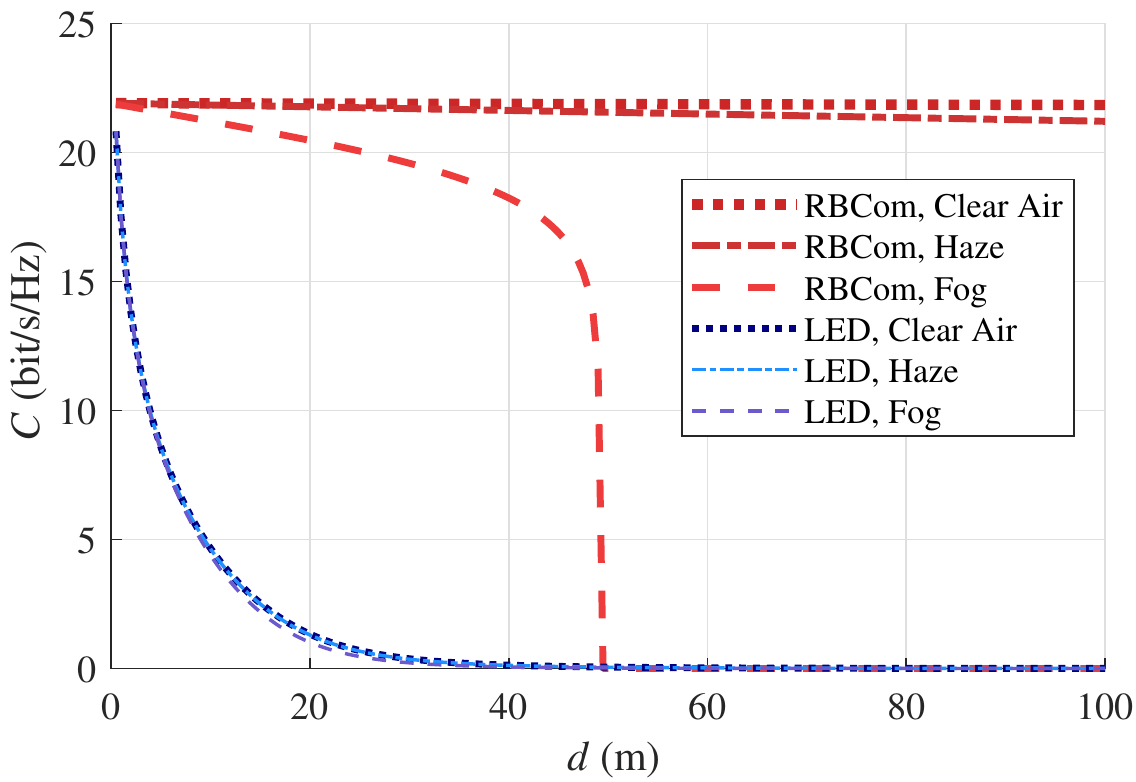}
	\caption{Comparison between RBCom system and LED-based system}
	\label{fig:CurvRbvsd}
\end{figure}

Next, we compare the performance of the RBCom system with that of the LED-based communication system. We only consider the case where the radiation angle and the incidence angle are zero. According to the LED-based communication channel model presented in~\cite{a180730.03,a180805.01}, the detected current is obtained as
\begin{equation}
I_{\rm sig,LED}=\dfrac{\rho \mathcal{A}_r(m+1)}{2\pi d^2}\dfrac{n^2}{\sin^2{\it \Psi}_c}T_s{\rm e}^{-\alpha d}P_{\rm t}
\end{equation}
where $\mathcal{A}_r$ is the receiving area, $m=-\ln2/\ln(\cos{\it \Phi}_{1/2})$ is the order of Lambertian emission, ${\it \Phi}_{1/2}$ is the semi-angle (at half power) of the LED, $n$ is the refractive index of a lens at the PD, ${\it\Psi}_c$ is the width of the field of vision at the receiver, $T_s$ is the gain of an optical filer, and $P_t$ is the transmit power. Thus, the SNR at the receiver of a LED-based system is 
\begin{equation}
	SNR_{\rm LED}=\dfrac{I_{\rm sig, LED}^2}{2q(I_{\rm sig,LED}+I_{\rm bk})B_{x}+4KTB_{x}/R_L}.
\end{equation}
Then, the channel capacity of the LED-based system can be obtained by~\eqref{equ:Capa}.
Here, we set $P_t=30$~W, $\mathcal{A}_r= 1$~cm$^2$, $n=1.5$, $T_s=1$, ${\it \Phi}_{1/2}=70^\circ$, ${\it \Psi}_c=60^\circ$~\cite{a180805.01}. We also assume that the radii of the elements of the RBCom system are great enough so that multiple-transverse-mode operation is assured. As shown in Fig.~\ref{fig:CurvRbvsd}, the RBCom channel capacity decreases slowly as the distance increases, but it decreases rapidly as the distance approaches the threshold. The   capacities of the RBCom channels are greater than that of the LED-based communication channel under different air conditions including clear air, haze, and fog.

\section{Conclusions}
\label{sec:con}
In this paper, the echo interference mechanism of the RBCom system was at first discussed. Then, the method of interference elimination was proposed, along with an exemplary design based on optical filtering and frequency shifting. We also presented the mathematic model and performance analysis of the exemplary system. Taking account for the impacts such as transverse mode distribution, phase variation, and pump power fluctuation is worthy of investigating for future work.




%

\setcounter{figure}{0}

\appendices

\section{Derivation of the Carrier Amplitude}
\label{apdx:carrieramp}

\renewcommand{\thefigure}{A.\arabic{figure}}

\begin{figure}
	\centering
	\includegraphics[width=3.3in]{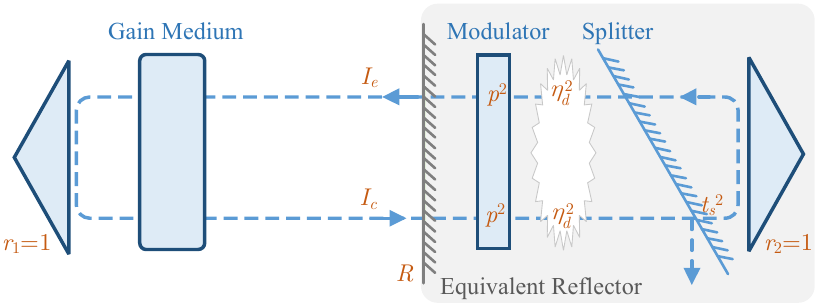}
	\caption{Equivalent model for computing the carrier amplitude}
	\label{fig:equvm}
\end{figure}

To obtain the carrier amplitude $A_c$, we need to compute the light intensity of the carrier $I_c$. Here, an equivalent model of the proposed interference-free RBCom system is depicted in Fig.~\ref{fig:equvm}. In this model, all the devices on the right-hand side of the gain medium are modeled as a partial reflector. Let $I_e$ denote the intensity of the reflected beam, the  intensity reflectivity of the equivalent reflector is derived as
\begin{equation}
	R=\dfrac{I_e}{I_c}=t_s^2 p^4\eta_d^4.
	\label{equ:R}
\end{equation}
In the equivalent model, the OBPF is neglected, as it imposes less attenuation on the carrier. Besides, we have made the assumption $r_1=r_2=1$ for this system. According to \eqref{equ:Gl-exam}, the DC bias $p$ which specifies the operating point of the modulator is regarded as the modulator's amplitude attenuation coefficient $\eta_m$. Hence, the two-pass intensity attenuation coefficient of the modulator is $p^4$, the two-pass intensity attenuation coefficient of transmission is $\eta_d^4$, and the intensity transmittance of the splitter is $t_s^2$.

The equivalent model is a classic laser cavity, and its output power is given by~\cite{a190702.01}
\begin{equation}
	P_{\rm out}=\dfrac{\mathcal{A}_b I_s\left[\left(g_0-\alpha_0\right)l+\ln{\sqrt{R}}\right]}{1-\dfrac{\alpha_0 l}{\ln\sqrt{R}}}
\end{equation}
where $\mathcal{A}_b$ is the beam cross section area (for multiple-transverse-mode operation, the cross section area of the laser beam is approximate to that of the gain medium), and $\alpha_0$ is the loss coefficient of the gain medium. Generally, $\alpha_0$ is very small compared with the small-signal gain factor $g_0$, so we set $\alpha_0=0$. As the intensity transmittance of the equivalent reflector is $(1-R)$, the carrier intensity is deduced as
\begin{equation}
	I_c=\dfrac{P_{\rm out}}{\mathcal{A}_b(1-R)}=\dfrac{I_s\left[g_0 l+\ln{\sqrt{R}}\right]}{1-R}.
\end{equation}
Therefore, the carrier amplitude is obtained as
\begin{equation}
A_c=\sqrt{Z_0I_c}=\sqrt{\dfrac{Z_0 I_s\left[g_0 l+\ln{\sqrt{R}}\right]}{1-R}}.
\label{equ:Ac-unsimp}
\end{equation}
The saturation intensity is given by~\cite{a181218.01}
\begin{equation}
I_s=\dfrac{hf_c}{\sigma \tau_f}
\label{equ:Is}
\end{equation}
where  $h$ is the Planck's constant, and $f_c$ is the carrier beam frequency.
The small-signal gain coefficient is given by~\cite{a181218.01}
\begin{equation}
	g_0=\dfrac{\sigma \tau_f\eta P_{\rm in}  }{h f_c V}=\dfrac{\eta P_{\rm in}}{I_sV}
\label{equ:g0}
\end{equation}
where $V$ is the volume of the gain medium. Here, we assume that the gain medium is a rod or disk with a cross section area $\mathcal{A}_b$ and a length $l$. Then we obtains
\begin{equation}
	g_0 l=\dfrac{\eta P_{\rm in}}{I_s\mathcal{A}_b}.
\label{equ:g0l}
\end{equation}
At last, substituting \eqref{equ:R}, \eqref{equ:Is}, and \eqref{equ:g0l} into \eqref{equ:Ac-unsimp} gives
\begin{equation}
A_c=\sqrt{\dfrac{Z_0 I_s\left[\dfrac{\eta P_{\rm in}}{I_s\mathcal{A}_b}+\ln{\sqrt{t_s^2 p^4\eta_d^4}}\right]}{1-t_s^2 p^4\eta_d^4}}.
\end{equation}

%



\ifCLASSOPTIONcaptionsoff
  \newpage
\fi




\bibliographystyle{IEEETran}
\small
%
\bibliography{mybib}
%
%

%

%
%







\end{document}